%% file: trustworthy-rag.tex
\documentclass[runningheads]{llncs}

\usepackage{url}
\PassOptionsToPackage{hyphens}{url}\usepackage{hyperref}
\usepackage{tabularx}
\usepackage{xcolor}
\usepackage{booktabs}
\usepackage[square,comma,numbers,sort&compress,sectionbib]{natbib}
\usepackage{enumerate}
\usepackage{graphicx}

\usepackage{fancyhdr}

\fancypagestyle{specialfooter}{%
  \fancyhf{}
  
  \fancyfoot[C]{\vspace{1cm} \hspace*{-.2\textwidth}\parbox{1.4\textwidth}{\small This is the author's version of the work. It is posted here for your personal use. The definitive version is published in: \\ \emph{Proceedings of the 48th European Conference on Information Retrieval} (ECIR '26), March 29--April 2, 2026, Delft, The Netherlands.}}
}

\begin{document}
\mainmatter

\titlerunning{Trust Me on This}
\title{Trust Me on This: A User Study of Trustworthiness for RAG Responses}

\author{Weronika Łajewska\thanks{Work done while at University of Stavanger, prior to joining Amazon.}\inst{1} \and
Krisztian Balog\inst{2}}
\authorrunning{Łajewska and Balog}

\tocauthor{Weronika Łajewska, Krisztian Balog}

\institute{Amazon \and University of Stavanger, Stavanger, Norway\\
\email{lajewska@amazon.lu, krisztian.balog@uis.no}}

\maketitle
\begin{abstract}
The integration of generative AI into information access systems often presents users with synthesized answers that lack transparency. This study investigates how different types of explanations can influence user trust in responses from retrieval-augmented generation systems. We conducted a controlled, two-stage user study where participants chose the more trustworthy response from a pair—one objectively higher quality than the other—both with and without one of three explanation types: (1) source attribution, (2) factual grounding, and (3) information coverage. Our results show that while explanations significantly guide users toward selecting higher quality responses, trust is not dictated by objective quality alone: Users’ judgments are also heavily influenced by response clarity, actionability, and their own prior knowledge.

\end{abstract}

\thispagestyle{specialfooter}

\input{trustworthy-rag-01}
\input{trustworthy-rag-02}
\input{trustworthy-rag-03}
\input{trustworthy-rag-04}
\input{trustworthy-rag-05}

\small{\subsubsection{\ackname} This research was partially supported by the Norwegian Research Center for AI Innovation, NorwAI (Research Council of Norway, project nr. 309834).}

\small{\subsubsection{Disclosure of Interests.}
The authors have no competing interests to declare that are relevant to the content of this article.}

\bibliographystyle{splncs04nat}
\bibliography{trustworthy-rag.bib}
\end{document}

%% file: trustworthy-rag-01.tex
\section{Introduction}

As generative AI becomes integrated into information access systems, from conversational agents to summaries on search engine result pages, users are increasingly presented with concise responses---often just a few sentences---without visibility into which sources the response was based on. 
This shift conceals crucial information that users typically rely on to assess the novelty, reliability, and relevance of retrieved content~\citep{Xu:2006:J.}.
Retrieval-augmented generation (RAG)~\citep{Lewis:2020:NIPS, Gienapp:2024:SIGIR, Izacard:2021:EACL, Ram:2023:Trans.} has recently emerged as a prominent approach to generating more factually grounded and diverse responses by leveraging external documents. However, RAG models still fall short on transparency: they offer no indication of low-confidence outputs or known limitations, whether stemming from incomplete retrieval or flaws in generation. Because users see only the final output, the responsibility lies with the system to surface potential issues, thereby promoting transparency and enabling users to judge the response quality.
Although explainability has been extensively studied in areas such as decision support and recommender systems~\citep{Balog:2020:SIGIR,Nunes:2017:User, Zhang:2020:FNT, Wang:2024:ToIS, Li:2023:ToIS, Wang:2021:IUI, Zhang:2020:FAT}, it remains underexplored in the context of retrieval-augmented response generation. 

While prior work establishes that explanations can enhance
the user-perceived \emph{usefulness} of a generated response~\citep{Lajewska:2024:SIGIR}, 
\emph{trust} operates on a different, more critical level.
Usefulness can be seen as the immediate utility of a response for a given task, relating to its clarity, format, or relevance. Trust, however, concerns the user's belief in the credibility and factual correctness of the information itself. A user might find a well-structured summary \emph{useful} for getting a quick overview, but they will only \emph{trust} it for a critical decision if they believe it to be factually sound. Therefore, trust is a distinct and vital prerequisite for the meaningful adoption of these systems, moving beyond surface-level utility to user confidence and reliance.
Therefore, this paper investigates the next logical step by asking: \emph{How do different types of explanations influence user trust in responses from retrieval-augmented generation systems?}

To answer this question, we conducted a controlled user study with a two-stage, within-subject design. We extended an existing response generation pipeline with a post-hoc module to generate three distinct types of explanations: (1) \emph{source attribution} via supporting passages, (2) \emph{factual grounding} by linking response statements to sources, and (3) \emph{information coverage} through the surfacing of relevant but omitted aspects of the topic.
Study participants were shown a pair of responses for a given query, where one response was of higher objective quality than the other. Initially, they chose the response they found more trustworthy. Subsequently, they were presented with the exact same pair of responses, but now enhanced with one of the explanation types, and were prompted to make their trust judgment again. This two-stage design enabled a direct measurement of how explanations altered a user's perception of trustworthiness.

The results of our study show that users’ trust judgments do not always align with the objective quality of responses. Many participants preferred the objectively lower quality response when it was clearer, more detailed, or more actionable. Source attribution had the strongest positive effect on trust, but only in factual or technical contexts, while it was largely ignored in subjective questions. Finally, as indicated in the free-text comments provided by users, the relevance of information and users’ own background knowledge strongly shaped trust decisions, with some participants dismissing or overlooking explanations when they felt confident in their own understanding of the topic. All collected user data, input responses, and corresponding annotations from this study are available in the public repository: \url{https://github.com/iai-group/trustworthy-rag/}.

%% file: trustworthy-rag-02.tex
\section{Related Work}

Trust is a key factor in human–machine interaction, particularly in information-intensive tasks such as retrieval-augmented response generation, which synthesizes relevant information into concise, comprehensive responses~\citep{Lewis:2020:NIPS}. 
System trustworthiness is not inherent but is instead communicated through trust cues---features of the interface, documentation, or explanations that shape users' judgments~\citep{Liao:2022:FAccT}. When these cues are misleading or poorly designed, they can foster misplaced trust or confusion~\citep{Liao:2022:FAccT}. Because such cues are often processed heuristically rather than analytically, users tend to rely on mental shortcuts that can result in inconsistent or biased trust assessments.
Providing explanations is one approach to addressing these challenges: explainable systems are expected to clarify their capabilities, communicate uncertainty, and make their reasoning transparent~\citep{Amershi:2019:CHI}. However, explanations can have mixed effects---they can enhance trust but may also lead to overtrust or inaccurate mental models, especially among users with low domain expertise~\citep{Cau:2023:IUI}.

Generated responses often contain unsupported claims or incorrect citations~\citep{Liu:2023:EMNLP}. Commercial conversational search systems, such as Perplexity.ai, Microsoft Copilot, and Google Gemini, display source links, but explanations are generally limited to attribution, overlooking other important aspects. 
Additional factors that influence trust include bias~\citep{Gao:2020:Inf, Draws:2021:SIGKDD}, unanswerability~\citep{Hu:2019:AAAI, Huang:2019:CoNLL}, and response completeness~\citep{Bolotova:2020:CIKM}. Prior work on trustworthy RAG and question answering has focused mainly on grounding and factual correctness (e.g., avoiding hallucinations) rather than directly on user trust~\citep{Li:2024:NAACL-HLT,Rashkin:2021:Comput.,Liu:2023:EMNLP, Schuster:2023:NAACL-HLT}.
High-quality explanations related to source, system confidence, and response limitations have been shown to improve the perceived usefulness of responses~\citep{Lajewska:2024:SIGIR}, echoing findings in broader AI research where calibrated confidence improves trust~\citep{Zhang:2020:FAT}. However, their effect on user trust remains largely unexplored. In this work, we adopt a user-centered perspective to investigate how different explanation strategies shape trust judgments in RAG responses for information-seeking queries.

%% file: trustworthy-rag-03.tex
\begin{figure}[t]
    \centering
    \includegraphics[width=0.8\linewidth]{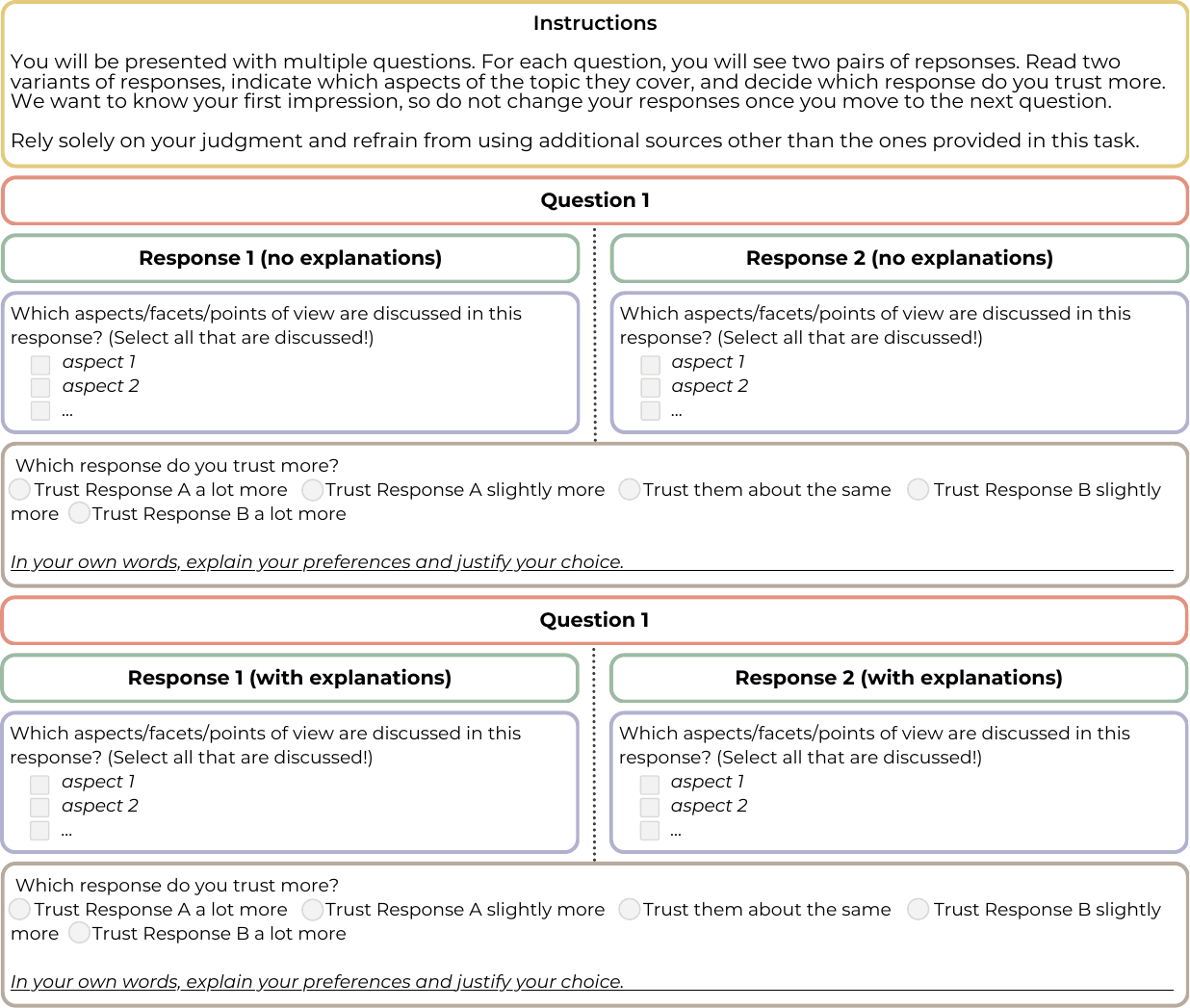}
    \caption{User study design (example on a single query).}
    \label{fig:study_design}
\end{figure}

\section{User Study}

To investigate the impact of different explanation types on users' trust in RAG responses, we designed and conducted a user study. The study evaluates three key dimensions of response quality---source attribution, factual grounding, and information coverage---each associated with a corresponding explanation type.
Our goal is to measure how truthful explanations revealing the quality of a response alter a user's initial trust judgment.

\subsubsection*{Experiment Design.}

We employed a within-subject study design where the core task involved a two-step comparison. As visualized in Figure \ref{fig:study_design}, participants were first presented with a pair of responses---one high-quality and one limited-quality, shown in random order---without explanations, and were asked the question: \emph{``Which response do you trust more?''}. 
After providing their initial judgment on a 5-point Likert scale, they were shown the exact same pair or responses, but this time enhanced with one of the explanation types. This design directly measures the shift in user trust attributable to the presence of an explanation.\footnote{Even though the modification of the scores for the first pair of responses (without explanations) is not blocked by the UI, users were instructed to not change it after moving to the next question.} 
After each comparison (both with and without explanations), users were also asked to justify their choice in a free-text comment.

We conducted separate experiments for each of the three explanation types, allowing us to isolate and evaluate the independent effect of each response dimension on user trust.

\subsubsection*{Responses and Explanations.}

We selected 30 queries from the TREC CAsT~'22 dataset~\citep{Owoicho:2022:TRECb} (10 for each response dimension). No more than two queries were selected from the same CAsT topic. Factual, information-intensive queries were prioritized for source attribution and factual grounding explanations, whereas queries involving complex or controversial topics were prioritized for information coverage explanations. For each query, responses were generated using GINGER, a modular pipeline for response creation based on information nuggets extracted from retrieved documents~\citep{Lajewska:2025:SIGIR}. This approach ensures grounding in specific facts, facilitates source attribution, and maximizes information inclusion within length constraints. 
To systematically study the effect of explanations, we manipulated the quality of the generated responses by altering the underlying source passages: 
\emph{High-quality responses} were based on relevant passages with high coverage of the information requested in the query.
\emph{Limited-quality responses} were based on only partially relevant passages covering a single aspect of the topic or were generated based solely on LLM's parametric memory without relying on external relevant passages---see Table~\ref{tab:explanation_types} for details. 

To generate explanations, we extended the GINGER pipeline with a post-hoc explanation module that includes source attribution, statement grounding, and additional topic facets revealment. 
Crucially, all explanations shown to users are truthful, accurately revealing the shortcomings of limited-quality responses.

\begin{table*}[t]
\centering
\scriptsize
\caption{Summary of explanation types and manipulated response characteristics. 
}
\label{tab:explanation_types}
\begin{tabular}{p{3.9cm}@{\hspace{8pt}}p{3.7cm}@{\hspace{8pt}}p{4cm}}
\toprule
\textbf{Response Dimension/} & \textbf{High-quality} & \textbf{Limited-quality} \\
\textbf{\textcolor{brown}{Explanation}} & \textbf{Response} & \textbf{Response} \\
\midrule
\textbf{Source Attribution} & Based on relevant passages with clear source attribution & Relying on information from LLM parametric memory \\
\textcolor{brown}{\textbf{Supporting passages}: Clarifies whether the response is grounded in verifiable sources} & \textcolor{brown}{\emph{List of supporting passages}} & \textcolor{brown}{\emph{``The response has no traceable origin and it cannot be attributed to specific source documents.''}} \\
\midrule
\textbf{Factual Grounding} & Each factual statement is traceable to a relevant passage & Response is generated from parametric memory without traceable origin \\
\textcolor{brown}{\textbf{Statement-level links}: Clarifies whether each claim in the response can be traced to a supporting source} & \textcolor{brown}{\emph{List of supporting passages linked inline to supported statements}} & \textcolor{brown}{\emph{``The response is not supported by verifiable references, and the individual claims cannot be reliably cross-checked against sources.''}} \\
\midrule
\textbf{Information Coverage} & Covers multiple facets of the query topic & Covers only a single aspect or misses major points \\
\textcolor{brown}{\textbf{Missing aspects highlighted}: Discloses whether important information is missing from the response} & \textcolor{brown}{\emph{``The response covers multiple aspects of the topic, providing a broad view.''}} & \textcolor{brown}{\emph{``The response focuses on just one aspect and may miss important points related to: Keyword: \{keyword\}''}} \\
\bottomrule
\end{tabular}
\end{table*}

\subsubsection*{Participants and Procedure.}

We recruited skilled and engaged crowd workers from Amazon Mechanical Turk through a qualification task containing 5 query-response pairs, each followed by a question about topic aspects covered in the given response. Each qualified worker completed a human intelligence task (HIT) that included 2 queries per explanation type, resulting in 6 unique queries per HIT. For each query, participants compared two response variants, first without explanations and then with explanations. In total, each worker evaluated 12 response comparisons. After each response comparison, workers were asked to justify their choice in a free-text comment. We follow a within-subject design to control for individual differences as each participant serves as their own control, reducing variability and increasing the sensitivity of the analysis, as well as to use recruited crowd workers more efficiently by collecting multiple data points from each one.
We designed 5 different HITs in total, each completed by 10 different participants.
This setup ensures balanced coverage and enables us to independently assess the effect of each type of explanation on user trust.

\subsubsection*{Study Execution.}

The qualification study and main data collection were conducted over a period of one week (18–25 September, 2025). The qualification task was released to 50 workers, of whom 36 met the threshold (correctly answering at least 4 out of 5 test questions) and were invited to participate in the user study. In total, 21 crowd workers contributed to the study: 6 completed all 5 HITs, 2 completed 3 HITs, 1 completed 2 HITs, and the remainder submitted a single HIT. Data quality was assessed through a control question accompanying each unique response, in which participants identified which aspects from a list were covered. Because a ground-truth set of aspects was available for each response, this served as a quality check. Participants correctly identified aspects in 425 out of 600 cases, indicating high overall data quality. The total cost of the study was \$316, including \$68 in Amazon Mechanical Turk fees. To further acknowledge high-quality contributions, four workers received a \$2 bonus for particularly insightful justifications of their trust decisions.

%% file: trustworthy-rag-04.tex
\section{Results and Analysis}

\begin{figure*}[t]
    \centering
    \includegraphics[width=\textwidth]{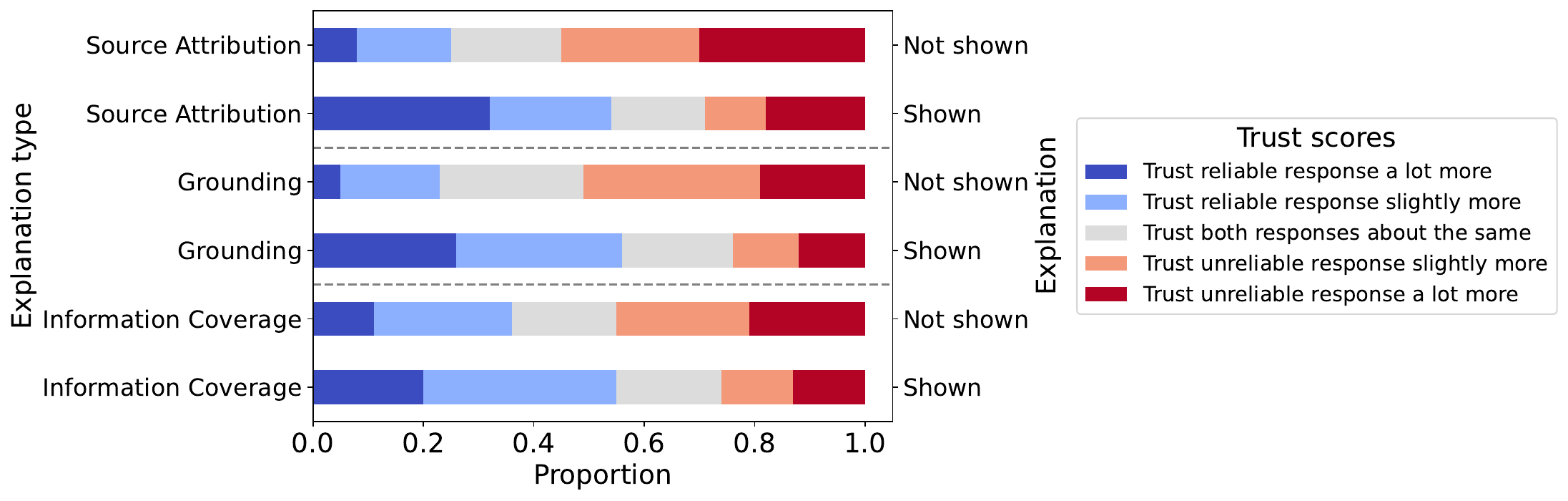}
    \caption{Proportion of user trust judgments for each explanation type, comparing responses with and without explanations. Colors indicate the direction of preference.}
    \label{fig:trust_scores_distribution}
\end{figure*}

Figure~\ref{fig:trust_scores_distribution} presents the distribution of trust preferences across explanation presence and types. Overall, showing explanations increases the likelihood of users selecting the reliable response, which was chosen as more trustworthy in 165/300 cases with explanations compared to 84/300 without. This result clearly indicates that providing explanations helps guide users toward more reliable information.

Delving into how users' perceptions change, we find that explanations prompt\-ed a shift in preference from the unreliable to the reliable response in 69 out of 300 cases. The effect varied by explanation type, with grounding causing the largest shift (27 cases), followed by source attribution (23 cases) and information coverage (19 cases).
This variation may also be influenced by the nature of queries: grounding queries were often factual and required information-intensive responses  \emph{(``I know that there are similar schemes to Ponzi schemes. Can you compare them?'')}. By contrast, for information coverage queries that are often opinion-based \emph{(``What do you think is the best Marvel movie?'')}, breadth and depth were easily assessed from the response alone. 

\subsubsection*{Qualitative Analysis.}

User justifications reveal a tension between objective quality and the perceived usability of responses. Participants often favored a limited-quality response when it was clearer or more actionable \emph{(``Response B gives practical, named resources ... which makes the advice more actionable'')}. Similarly, specific examples frequently boosted perceived trustworthiness regardless of their factuality \emph{(``Response A provides clear examples of geographic and economic biases ... giving a broad perspective'')}.

Information coverage also emerged as a recurring theme, though its effects were mixed. Participants valued comprehensive coverage when it was relevant \emph{(``Response B covers a broader range including the resignation of aides, which adds depth to the impact'')}, but viewed it as distracting when it did not directly address the query \emph{(``Response B’s focus on a broader view isn’t necessarily helpful, given the query is specifically asking for meat dishes'')}.

Source attribution was another important factor, though its influence was highly context-dependent. In fact-based domains, sources often tipped the balance \emph{(``Without citations, response B’s numbers can’t be verified, while response A’s quoted sources look authoritative'')}. Yet in subjective queries, participants tended to dismiss their importance \emph{(``Sources are not needed for subjective questions''; ``I don’t care about response B’s citing of sources, since this is an inherently subjective question'')}. While preliminary, these findings highlight the need for further investigation into how explanation strategies vary across query types.

Finally, users’ background knowledge shaped trust decisions when they could not easily assess response quality. In such cases, participants either trusted both responses equally \emph{(``Without any previous knowledge, there isn’t much that I can really say here. No real difference'')} or relied on their own expertise to guide judgments \emph{(``... I know for sure response one is factually correct but response two cannot be verified, which makes me even discredit the response more'')}. Since users’ background knowledge was not explicitly measured in this study, these observations are indicative rather than conclusive and warrant a more systematic and focused analysis of the role of background knowledge in future work.

%% file: trustworthy-rag-05.tex
\section{Conclusion}

We conducted a user study to examine how the presence and type of explanations affect user trust in responses from retrieval-augmented generation systems. 
Our findings show that users’ trust judgments often diverge from the objective quality of responses, with many favoring limited-quality responses that seem clearer, more detailed, or more actionable. 
Based on the results of our study, future response generation systems should adapt explanations based on query type and response complexity, use source-based explanations primarily for factual domains, and personalize explanations according to users’ prior knowledge.
Future work should also investigate the long-term effects of these explanations on how users' trust evolves and calibrates over time.

%% file: trustworthy-rag.bbl
\begin{thebibliography}{27}
\providecommand{\natexlab}[1]{#1}
\providecommand{\url}[1]{\texttt{#1}}
\providecommand{\urlprefix}{URL }
\expandafter\ifx\csname urlstyle\endcsname\relax
  \providecommand{\doi}[1]{doi:\discretionary{}{}{}#1}\else
  \providecommand{\doi}{doi:\discretionary{}{}{}\begingroup
  \urlstyle{rm}\Url}\fi

\bibitem[{Amershi et~al.(2019)Amershi, Weld, Vorvoreanu, Fourney, Nushi,
  Collisson, Suh, Iqbal, Bennett, Inkpen, Teevan, {Kikin-Gil}, and
  Horvitz}]{Amershi:2019:CHI}
Amershi, S., Weld, D., Vorvoreanu, M., Fourney, A., Nushi, B., Collisson, P.,
  Suh, J., Iqbal, S., Bennett, P.N., Inkpen, K., Teevan, J., {Kikin-Gil}, R.,
  Horvitz, E.: Guidelines for human-{AI} interaction. In: Proceedings of the
  2019 CHI Conference on Human Factors in Computing Systems, pp. 1--13, CHI '19
  (2019)

\bibitem[{Balog and Radlinski(2020)}]{Balog:2020:SIGIR}
Balog, K., Radlinski, F.: Measuring recommendation explanation quality: The
  conflicting goals of explanations. In: Proceedings of the 43rd International
  ACM SIGIR Conference on Research and Development in Information Retrieval,
  pp. 329--338, SIGIR '20 (2020)

\bibitem[{Bolotova et~al.(2020)Bolotova, Blinov, Zheng, Croft, Scholer, and
  Sanderson}]{Bolotova:2020:CIKM}
Bolotova, V., Blinov, V., Zheng, Y., Croft, W.B., Scholer, F., Sanderson, M.:
  Do people and neural nets pay attention to the same words: Studying
  eye-tracking data for non-factoid {QA} evaluation. In: Proceedings of the
  29th ACM International Conference on Information \& Knowledge Management, pp.
  85--94, CIKM '20 (2020)

\bibitem[{Cau et~al.(2023)Cau, Hauptmann, Spano, and Tintarev}]{Cau:2023:IUI}
Cau, F.M., Hauptmann, H., Spano, L.D., Tintarev, N.: Supporting
  high-uncertainty decisions through {AI} and logic-style explanations. In:
  Proceedings of the 28th International Conference on Intelligent User
  Interfaces, pp. 251--263, IUI '23 (2023)

\bibitem[{Draws et~al.(2021)Draws, Tintarev, and Gadiraju}]{Draws:2021:SIGKDD}
Draws, T., Tintarev, N., Gadiraju, U.: Assessing viewpoint diversity in search
  results using ranking fairness metrics. ACM SIGKDD Explorations Newsletter
  \textbf{23}(1), 50--58 (2021)

\bibitem[{Gao and Shah(2020)}]{Gao:2020:Inf}
Gao, R., Shah, C.: Toward creating a fairer ranking in search engine results.
  Information Processing \& Management \textbf{57}(1), 102--138 (2020)

\bibitem[{Gienapp et~al.(2024)Gienapp, Scells, Deckers, Bevendorff, Wang,
  Kiesel, Syed, Fr\"{o}be, Zuccon, Stein, Hagen, and
  Potthast}]{Gienapp:2024:SIGIR}
Gienapp, L., Scells, H., Deckers, N., Bevendorff, J., Wang, S., Kiesel, J.,
  Syed, S., Fr\"{o}be, M., Zuccon, G., Stein, B., Hagen, M., Potthast, M.:
  Evaluating generative ad hoc information retrieval. In: Proceedings of the
  47th International ACM SIGIR Conference on Research and Development in
  Information Retrieval, p. 1916–1929, SIGIR '24 (2024)

\bibitem[{Hu et~al.(2019)Hu, Wei, Peng, Huang, Yang, and Li}]{Hu:2019:AAAI}
Hu, M., Wei, F., Peng, Y., Huang, Z., Yang, N., Li, D.: Read + verify: Machine
  reading comprehension with unanswerable questions. In: Proceedings of the
  AAAI Conference on Artificial Intelligence, pp. 6529--6537, AAAI '19 (2019)

\bibitem[{Huang et~al.(2019)Huang, Tang, Huang, He, and
  Zhou}]{Huang:2019:CoNLL}
Huang, K., Tang, Y., Huang, J., He, X., Zhou, B.: Relation module for
  non-answerable predictions on reading comprehension. In: Proceedings of the
  23rd Conference on Computational Natural Language Learning, pp. 747--756,
  CoNLL '19 (2019)

\bibitem[{Izacard and Grave(2021)}]{Izacard:2021:EACL}
Izacard, G., Grave, E.: Leveraging passage retrieval with generative models for
  open domain question answering. In: Proceedings of the 16th Conference of the
  European Chapter of the Association for Computational Linguistics: Main
  Volume, pp. 874--880, EACL '21 (2021)

\bibitem[{{\L}ajewska and Balog(2025)}]{Lajewska:2025:SIGIR}
{\L}ajewska, W., Balog, K.: {GINGER}: Grounded information nugget-based
  generation of responses. In: Proceedings of the 48th International ACM SIGIR
  Conference on Research and Development in Information Retrieval, p.
  2723–2727, SIGIR '25 (2025)

\bibitem[{{\L}ajewska et~al.(2024){\L}ajewska, Spina, Trippas, and
  Balog}]{Lajewska:2024:SIGIR}
{\L}ajewska, W., Spina, D., Trippas, J., Balog, K.: Explainability for
  transparent conversational information-seeking. In: Proceedings of the 47th
  International ACM SIGIR Conference on Research and Development in Information
  Retrieval, pp. 1040--1050, SIGIR '24 (2024)

\bibitem[{Lewis et~al.(2020)Lewis, Perez, Piktus, Petroni, Karpukhin, Goyal,
  K{\"u}ttler, Lewis, Yih, Rockt{\"a}schel, Riedel, and
  Kiela}]{Lewis:2020:NIPS}
Lewis, P., Perez, E., Piktus, A., Petroni, F., Karpukhin, V., Goyal, N.,
  K{\"u}ttler, H., Lewis, M., Yih, W.t., Rockt{\"a}schel, T., Riedel, S.,
  Kiela, D.: Retrieval-augmented generation for knowledge-intensive {NLP}
  tasks. In: Proceedings of the 34th International Conference on Neural
  Information Processing Systems, pp. 9459--9474, NIPS '20 (2020)

\bibitem[{Li et~al.(2023)Li, Zhang, and Chen}]{Li:2023:ToIS}
Li, L., Zhang, Y., Chen, L.: Personalized prompt learning for explainable
  recommendation. ACM Transactions on Information Systems \textbf{41}(4), 1--26
  (2023)

\bibitem[{Li et~al.(2024)Li, Park, Lee, and Bastani}]{Li:2024:NAACL-HLT}
Li, S., Park, S., Lee, I., Bastani, O.: {TRAQ}: Trustworthy retrieval augmented
  question answering via conformal prediction. In: Duh, K., Gomez, H., Bethard,
  S. (eds.) Proceedings of the 2024 Conference of the North American Chapter of
  the Association for Computational Linguistics: Human Language Technologies
  (Volume 1: Long Papers), NAACL-HLT '24 (2024)

\bibitem[{Liao and Sundar(2022)}]{Liao:2022:FAccT}
Liao, Q., Sundar, S.S.: Designing for responsible trust in {AI} systems: A
  communication perspective. In: Proceedings of the 2022 ACM Conference on
  Fairness, Accountability, and Transparency, pp. 1257--1268, FAccT '22 (2022)

\bibitem[{Liu et~al.(2023)Liu, Zhang, and Liang}]{Liu:2023:EMNLP}
Liu, N., Zhang, T., Liang, P.: Evaluating verifiability in generative search
  engines. In: Findings of the Association for Computational Linguistics: EMNLP
  2023, pp. 7001--7025, EMNLP '23 (2023)

\bibitem[{Nunes and Jannach(2017)}]{Nunes:2017:User}
Nunes, I., Jannach, D.: A systematic review and taxonomy of explanations in
  decision support and recommender systems. User Modeling and User-Adapted
  Interaction \textbf{27}(3-5), 393--444 (2017)

\bibitem[{Owoicho et~al.(2022)Owoicho, Dalton, Aliannejadi, Azzopardi, Trippas,
  and Vakulenko}]{Owoicho:2022:TRECb}
Owoicho, P., Dalton, J., Aliannejadi, M., Azzopardi, L., Trippas, J.R.,
  Vakulenko, S.: {TREC} {CAsT} 2022: Going beyond user ask and system retrieve
  with initiative and response generation. In: The Thirty-First Text REtrieval
  Conference Proceedings, TREC '22 (2022)

\bibitem[{Ram et~al.(2023)Ram, Levine, Dalmedigos, Muhlgay, Shashua,
  {Leyton-Brown}, and Shoham}]{Ram:2023:Trans.}
Ram, O., Levine, Y., Dalmedigos, I., Muhlgay, D., Shashua, A., {Leyton-Brown},
  K., Shoham, Y.: In-context retrieval-augmented language models. Transactions
  of the Association for Computational Linguistics \textbf{11}, 1316--1331
  (2023)

\bibitem[{Rashkin et~al.(2021)Rashkin, Nikolaev, Lamm, Aroyo, and
  Collins}]{Rashkin:2021:Comput.}
Rashkin, H., Nikolaev, V., Lamm, M., Aroyo, L., Collins, M.: Measuring
  attribution in natural language generation models. Computational Linguistics
  \textbf{49}(4), 777--840 (2021)

\bibitem[{Schuster et~al.(2023)Schuster, Lelkes, Sun, Gupta, Berant, Cohen, and
  Metzler}]{Schuster:2023:NAACL-HLT}
Schuster, T., Lelkes, A.D., Sun, H., Gupta, J., Berant, J., Cohen, W.W.,
  Metzler, D.: {SEMQA}: Semi-extractive multi-source question answering. In:
  Proceedings of the 2024 Conference of the North American Chapter of the
  Association for Computational Linguistics: Human Language Technologies, pp.
  1363--1381, NAACL-HLT '23 (2023)

\bibitem[{Wang et~al.(2024)Wang, Zhang, Wang, and Ricci}]{Wang:2024:ToIS}
Wang, S., Zhang, X., Wang, Y., Ricci, F.: Trustworthy recommender systems. ACM
  Transactions on Intelligent Systems and Technology \textbf{15}(4), 1--20
  (2024)

\bibitem[{Wang and Yin(2021)}]{Wang:2021:IUI}
Wang, X., Yin, M.: Are explanations helpful? a comparative study of the effects
  of explanations in {AI}-assisted decision-making. In: Proceedings of the 26th
  International Conference on Intelligent User Interfaces, pp. 318--328, IUI
  '21' (2021)

\bibitem[{Xu and Chen(2006)}]{Xu:2006:J.}
Xu, Y.C., Chen, Z.: Relevance judgment: What do information users consider
  beyond topicality? Journal of the American Society for Information Science
  and Technology \textbf{57}(7), 961–973 (2006)

\bibitem[{Zhang and Chen(2020)}]{Zhang:2020:FNT}
Zhang, Y., Chen, X.: Explainable recommendation: A survey and new perspectives.
  Foundations and Trends{\textregistered} in Information Retrieval
  \textbf{14}(1), 1--101 (2020)

\bibitem[{Zhang et~al.(2020)Zhang, Liao, and Bellamy}]{Zhang:2020:FAT}
Zhang, Y., Liao, Q.V., Bellamy, R.K.E.: Effect of confidence and explanation on
  accuracy and trust calibration in {AI}-assisted decision making. In:
  Proceedings of the 2020 Conference on Fairness, Accountability, and
  Transparency, pp. 295--305, FAT '20 (2020)

\end{thebibliography}
